\documentclass[prb,preprint]{revtex4-1} 
\usepackage{amsmath}  % needed for \tfrac, \bmatrix, etc.
\usepackage{amsfonts} 
\usepackage{graphicx} % needed for figures
\usepackage{subcaption}
\begin{document}
	
	\title{Nonlinear Coupling between Magnetic Gears}
	
	\date{\today}
	\author{Tianchi Liu}
	\affiliation{School of Physics, Nanjing University, Nanjing 210093,
	People's Republic of China}
	\email{231840003@smail.nju.edu.cn}
	
	\begin{abstract}
		This study investigates the complex nonlinear coupling of magnetic gears arranged in proximity on a plane. Acknowledging the rich array of geometric and electromagnetic parameters involved, we initiate our exploration with a simplified model. By nondimensionalizing the key variables, we derive a novel nonlinear dynamics framework that abstracts away electromagnetic dependencies. Our approach includes analatic results, numerical simulations and experimental validation, emphasizing two distinct operational modes: free motion and constant driving motion. The findings highlight the intricate behaviors arising from these interactions, contributing to a deeper understanding of magnetic gear dynamics.
	\end{abstract}
	
	\maketitle
	
	\section{Introduction}
	A magnetic gear resembles the traditional mechanical gear in geometry and function, using magnets instead of teeth. As two opposing magnets approach each other, they repel; when placed on two rings the magnets will act like teeth. Despite their practical importance, the nonlinear dynamics arising when these gears are simply aligned side by side remain underexplored.
	
	In this article, we'll introduce a simple model, which is a simple spur gear consists of N pairs of magnetic dipoles.
	A simple experimental setup can be achieved by placing cylindrical neodymium magnets at the ends of several fidget spinners. While it is worth mentioning that the directions of every magnetic moment could be either up, down, or even aligned with the arms of the spinners, it is desirable to retain the rotational symmetry of the system to a certain extent. Therefore, we categorize the arrangement of the magnets in two aspects: alignment and configuration. Alignment refers to the direction of all magnetic moments in the system, which can be along the $z$-axis or radial. Common configurations include fully aligned, mirror-image, and staggered arrangements. We focused our attention on staggered arrangements with dipoles along with $z$-axis.
	
	First, this article will present the analytical solution when magnetic gears are released at a small angle or velocity. The nonlinear system becomes linear due to small-angle approximation. Next,we will present the poincare sections for different energy state and initial conditions. Then, we will use potential-map to help us analysis the free motion behavior, which is a highlight of our research. Finally, we will cast a light on oscillation during stable transimission, which is a main factor that the spur gears are less widely used than coaxial magnetic gears.\cite{spur}\cite{slip}

	Nonlinear systems exhibit a variety of phenomena due to complex interactions, often influenced by restoring forces. A classic example is the double pendulum, which demonstrates nonlinear behavior such as sensitive dependence on initial conditions, leading to unpredictable and chaotic motion. This system displays a range of behaviors including irregular oscillations and complex bifurcations. Additionally, nonlinear damping effects have been an area of active study in recent years. To facilitate learning, various experimental devices have been developed for educational settings, allowing students to observe nonlinear dynamics in action. 
	
	This problem has aroused extensive interest among undergraduates and even high school students as a popular competition problem in the 2024 International Young Physicists’ Tournament and the 2024 China Undergraduate Physics Tournament. It also provides an example of the phenomenon of nonlinear oscillators and chaos. The phenomenon can easily be reproduced experimentally in a student laboratory. It provides a perfect example to demonstrate, within the scope of undergraduate physics, the rich dynamics associated with the effects of nonlinear free motion or driving.  
	
	\section{Theoretical Model}
	\subsection{Dynamic Equations}
	Consider two spinners with $N$ pairs of magnetic dipoles attached at their ends. The system has two degrees of freedom, and the state equations are determined by the coordinates $\theta_1$, $\theta_2$, $\omega_1$, $\omega_2$. The kinetic energy of the system can be described as:
	\begin{equation}
		T = \frac{1}{2}I \omega_1^2 + \frac{1}{2}I \omega_2^2,
	\end{equation} 	
	where $I$ is the moment of inertia of each magnetic gear.
	
	To determine the interaction potential between them, we need only sum up the potential of each pair of dipoles individually. We can write this in dimensionless form:
	\begin{equation}
		V \bigg/ \left( \frac{\mu_0 m^2}{4 \pi R^{3}} \right) = \sum_{i=1}^{N} \sum_{j=1}^{N} -\frac{\hat{e}_i \cdot \hat{e}_j}{|\vec{r}_{ij}|^3},
	\end{equation}
	where $m$ is the magnitude of the dipole moment, $R$ is the distance between the dipole and its spinning axis, $\hat{e}_i$, $\hat{e}_j$ represent the directions of the moments, and $\vec{r}_{ij}$ is the vector between two dipoles, which can be expressed as:
	\begin{equation}
		\vec{r}_{12} = - \hat{r}_{1} + \xi \hat{e}_{x} + \hat{r}_{2}.
	\end{equation}
	Here, $\xi$ is a parameter representing the ratio of the distance $d$ between the two spinners and $R$. We can then write the dynamical equations in dimensionless form as:
	\begin{eqnarray}
		\frac{d^2 \theta_1}{d \tau^2} = -\frac{\partial V}{\partial \theta_1}, \\
		\frac{d^2 \theta_2}{d \tau^2} = -\frac{\partial V}{\partial \theta_2},
	\end{eqnarray}
	where $\tau$ is defined as
	\begin{equation}
		\tau = \omega_0 t,\quad \omega_0 = \sqrt{\frac{\mu_0 m^2}{4 \pi R^{3} I}}.
	\end{equation}
	
	\subsection{Harmonic Oscillation}
\begin{figure}[h!]
		\centering
		\includegraphics[width=\textwidth]{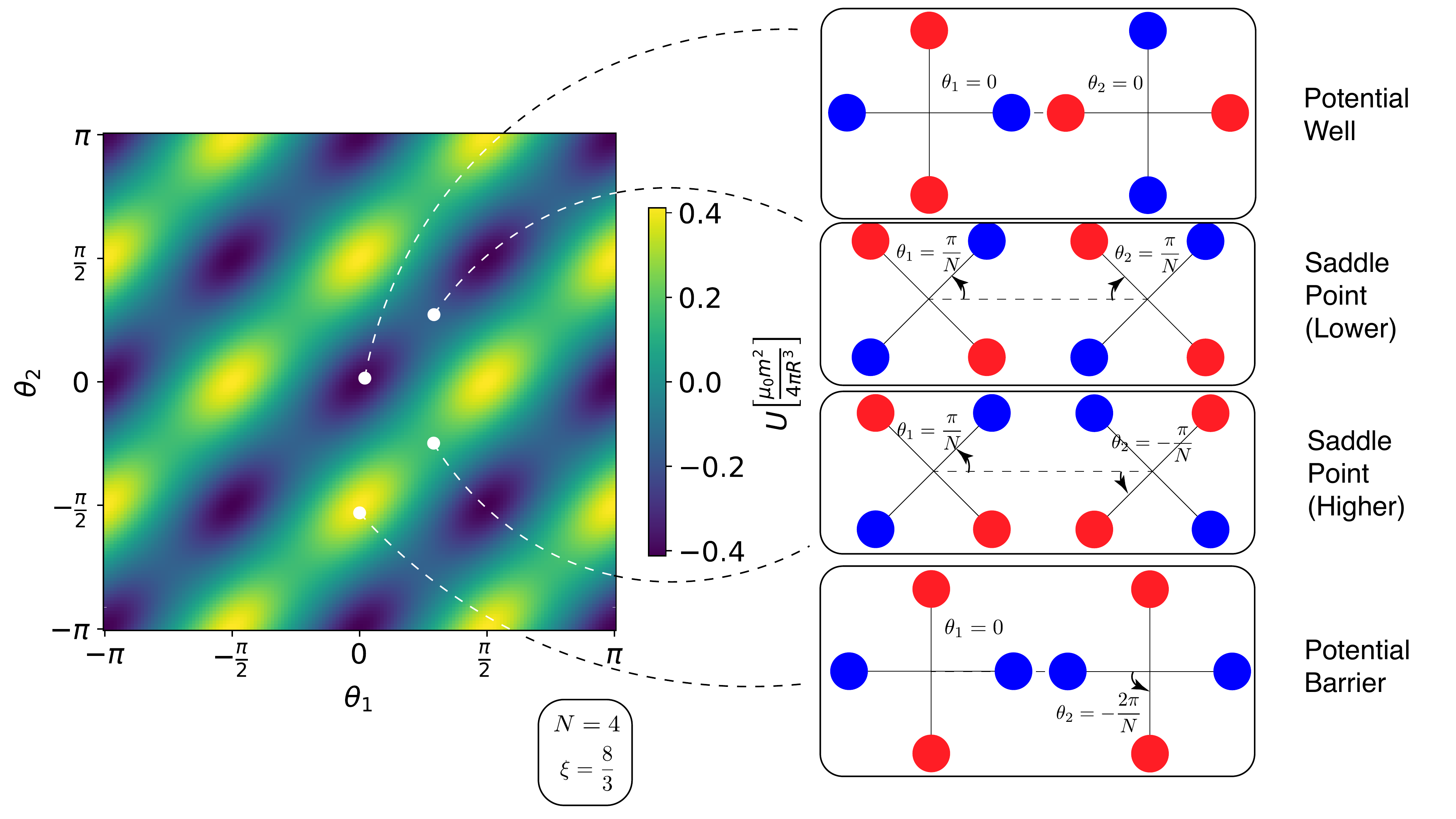}
		\caption{There are four characteristic potential points.}
		\label{fig:diagram}
	\end{figure}

	In the following, we consider the simplest form of the magnetic gear system: two spinners with only one dipole attached to each end. Furthermore, their dipole moments are aligned along the $z$-axis, and the configuration follows a mirror-image arrangement. The coordinate system is chosen as shown in the 	\ref{fig:diagram}. The two dipoles attract each other, and the system reaches an equilibrium point when the two dipoles are closest to each other (i.e., $\theta_1 = \theta_2 = 0$). Now the potential can be expanded around the equilibrium point as follows:
	\begin{equation}
		V = \frac{3}{(\xi - 2)^4} \left( \frac{\theta_1 + \theta_2}{2} \right)^2 + \frac{3 \xi}{(\xi - 2)^5} \left( \frac{\theta_1 - \theta_2}{2} \right)^2 + O(\theta_1^3, \theta_2^3).
	\end{equation}
	By introducing new normal modes, the potential energy function can be rewritten in a more compact form:
	\begin{equation}
		V = \frac{3}{(\xi - 2)^4} \bar{\theta}^2 + \frac{3 \xi}{(\xi - 2)^5} \Delta \theta^2 + O(\theta_1^3, \theta_2^3),
	\end{equation}
	where $\bar{\theta} = \frac{\theta_1 + \theta_2}{2}$ and $\Delta \theta = \frac{\theta_1 - \theta_2}{2}$. This reveals that when the system's energy is sufficiently low, the motion of the two gears becomes a combination of two simple harmonic motions, where the gears either rotate in the same direction or in opposite directions. We can always choose a suitable parameter $\xi$ such that the ratio of the periods of the two motions is a rational number, thus presenting a Lissajous figure	\ref{fig:lissajous_a}. In an ideal, non-dissipative system, the Hamiltonian is conserved, and it takes a particularly simple form under small oscillations\ref{fig:lissajous_b}:
		\begin{figure}[h!]
		\centering
		\begin{subfigure}[b]{0.45\textwidth}
			\centering
			\includegraphics[width=\textwidth]{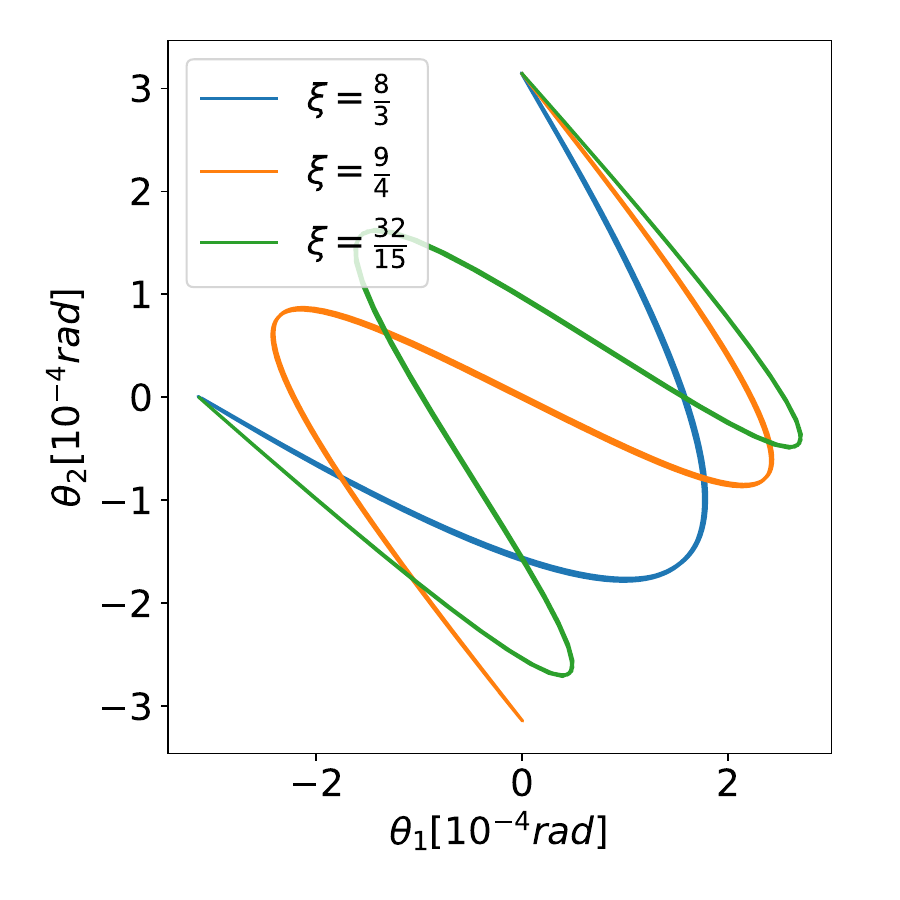}
			\caption{}
			\label{fig:lissajous_a}
		\end{subfigure}
		\hfill
		\begin{subfigure}[b]{0.45\textwidth}
			\centering
			\includegraphics[width=\textwidth]{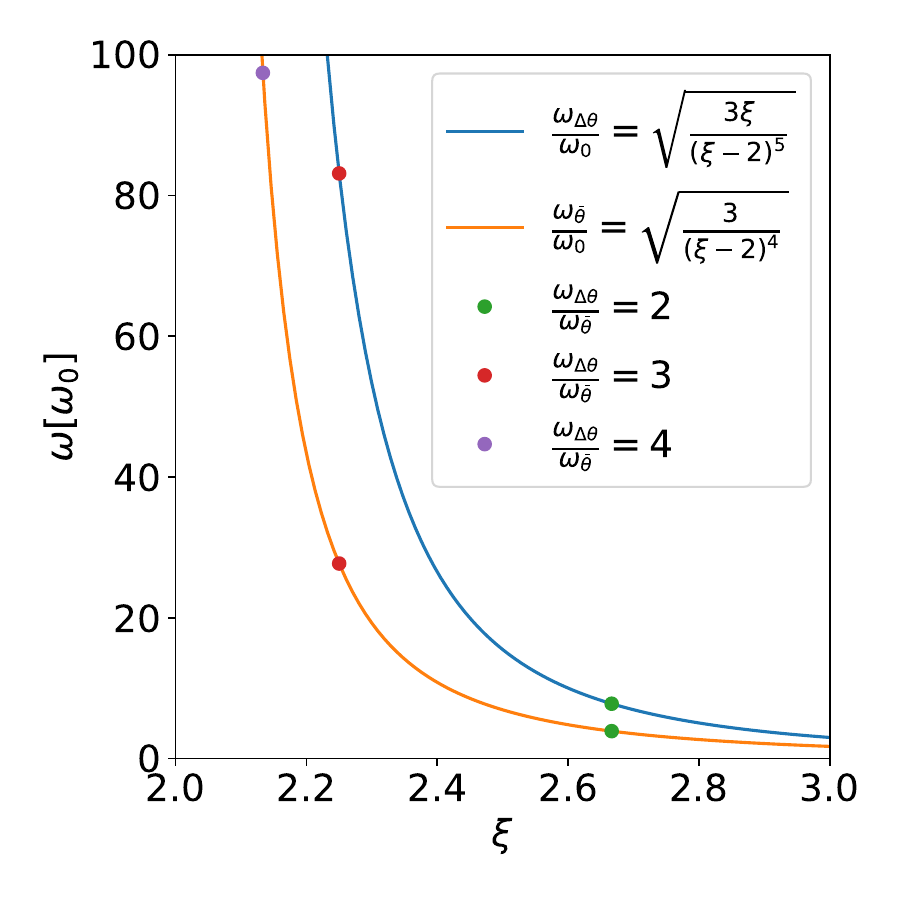}
			\caption{}
			\label{fig:lissajous_b}
		\end{subfigure}
		\caption{(a) Lissajous figure illustrating the motion of the gears. (b) Frequency ratio of the normal modes.}
	\end{figure}
	
	\begin{equation} 		
		H = \left( \frac{d \bar{\theta}}{d \tau} \right)^2 + \left( \frac{d \Delta \theta}{d \tau} \right)^2 + \frac{3}{(\xi - 2)^4} \bar{\theta}^2 + \frac{3 \xi}{(\xi - 2)^5} \Delta \theta^2.
	\end{equation}

	\subsection{Nonlinear Behavior}
\begin{figure}[h!]
		\centering
		\includegraphics[width=\textwidth]{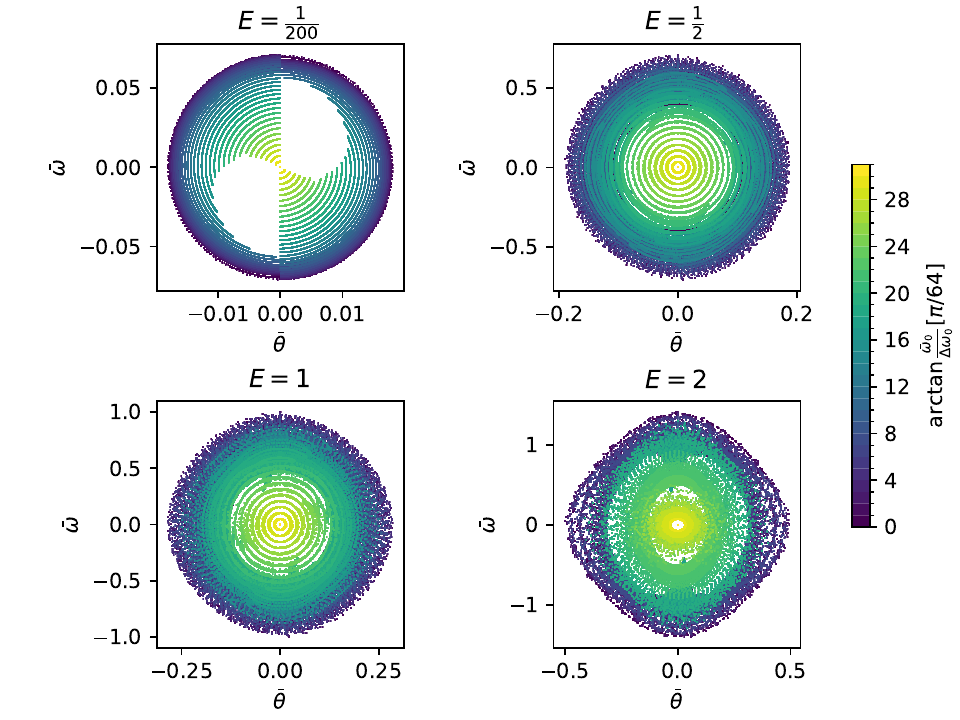}
		\caption{Poincaré section for different initial energy states (2500 time units).}
		\label{fig:poincare}
	\end{figure}

	The discussion above is taken under the small oscillation thesis, not enough to explain the phenomenon when the system behave as a cogwheel. We gradually increase the initial velocity(i.e. the energy) and track the trajectory. 
	However, the system has four coordinates, which are $\theta_1$, $\theta_2$,$\omega_1$, $\omega_2$, making it hard to be drawn on a picture. We use 'poincare map' to help analysis. 
	Every time the state crosses the plane $\omega_1-\omega_2=0$, we draw a point.\ref{fig:poincare} Generally, it shows an Elliptic fixed point at the center, and the trajectory gradually evolves in the four-dimensional space. As the increase of the energy state, the cross section is contorted from an elliptic shape. Different trajectory are becoming overlapped with each other. When the energy of the system reaches a threshold, the 'particle' will escape from the fixed point i.e. the potential well and could hardly cross the plane $\omega_1-\omega_2=0$. 
	
	\section{Potential Map}
	\subsection{Slippage}
	When the two opposite dipoles are closest to each other, the system's potential energy reaches its lowest state. Every time the second spinner rotates $2\pi$, the system state is recovered. Now we consider the general situation where each spinner has $N$ staggered dipoles on it. Naturally, the spinners will come to an equilibrium point similar to the one-dipole-per-spinner system. The simplest way for the system to recover is to rotate by an angle of $\frac{\pi}{N}$ each, as shown in the figure. A barrier will be encountered along the way (figure needed). On the potential map, it is the lower saddle point, and we call it the \textit{barrier point} to distinguish it from the higher saddle point.
	
	If the second spinner rotates really fast by an angle of $\frac{\pi}{N}$, the system's potential will suddenly increase because the leading term (i.e., the closest two dipoles) are no longer attracting each other but repelling. The first gear tends to rotate in the same direction, as if the two cogwheels are engaged. As the second gear keeps moving another angle of $\frac{\pi}{N}$, the system state is recovered. We call this phenomenon \textit{slippage}. But it costs a lot of energy to get to another potential trap as it crosses the highest hill. More commonly, the two spinners will rotate in the opposite direction, crossing the highest saddle point and coming to another potential trap, moving antidiagonally as it behaves on the potential map. This is not desired because this kind of dislocation will eventually make the system chaotic, especially if we take the damping term into consideration. We ask the question: how can we make the dislocation less likely to happen?
	
	\subsection{Free Motion Dynamical Behavior}
	We plot the heights of three characteristic points on the same figure. If the initial kinetic energy is even lower than the first saddle point, the imagined particle on the potential map will oscillate nonlinearly in the trap. If the energy is between the two saddle points, it will either bypass the barrier diagonally or antidiagonally to get to the next trap. Even if the energy could be higher than the barrier, it is quite critical for it to reach the highest point. It's always cheaper to bypass the barrier.
	
	But what if the second saddle point's height could be distinguished from the barrier? Naturally, the imagined particle tends to choose the cheapest path: diagonally (i.e., both spinners moving together). If the initial condition pushes the particle to move antidiagonally, it will somehow gain some diagonal speed and lose some in the opposite direction. The mechanism behind this is that any disturbance (not exactly antidiagonal) could be magnified in this nonlinear system. It will be scattered until it reaches a balance in a single diagonal trap. Now it behaves as a cogwheel.
	\begin{figure}[h!]
		\centering
		\begin{subfigure}[b]{0.45\textwidth}
			\centering
			\includegraphics[width=\textwidth]{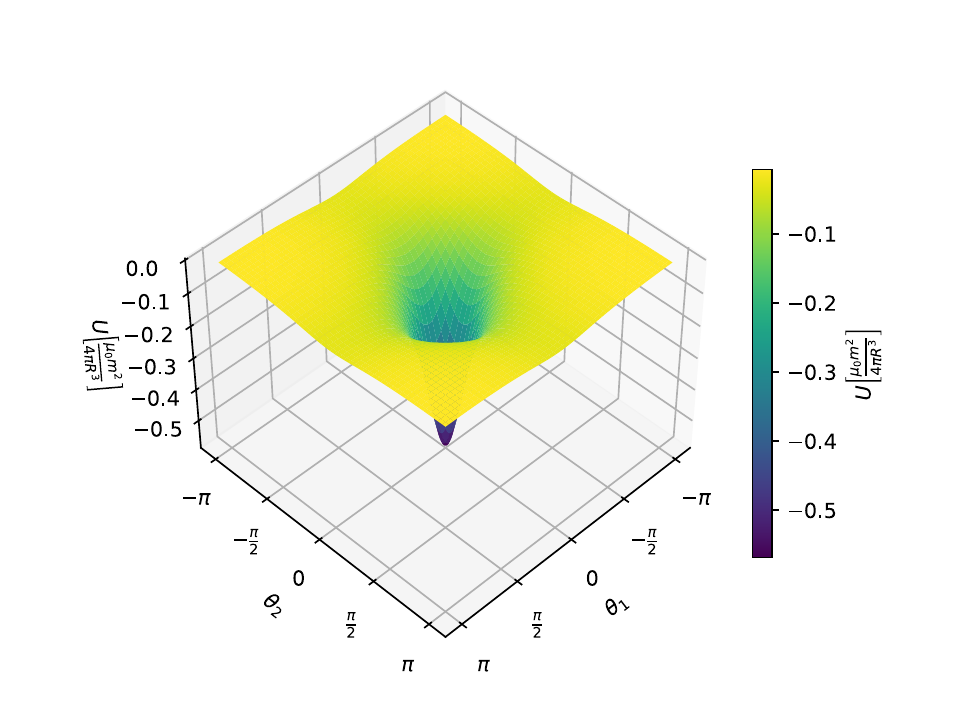}
			\caption{$N = 1$}
			\label{fig:potential_plot_3D_N_1}
		\end{subfigure}
		\hfill
		\begin{subfigure}[b]{0.45\textwidth}
			\centering
			\includegraphics[width=\textwidth]{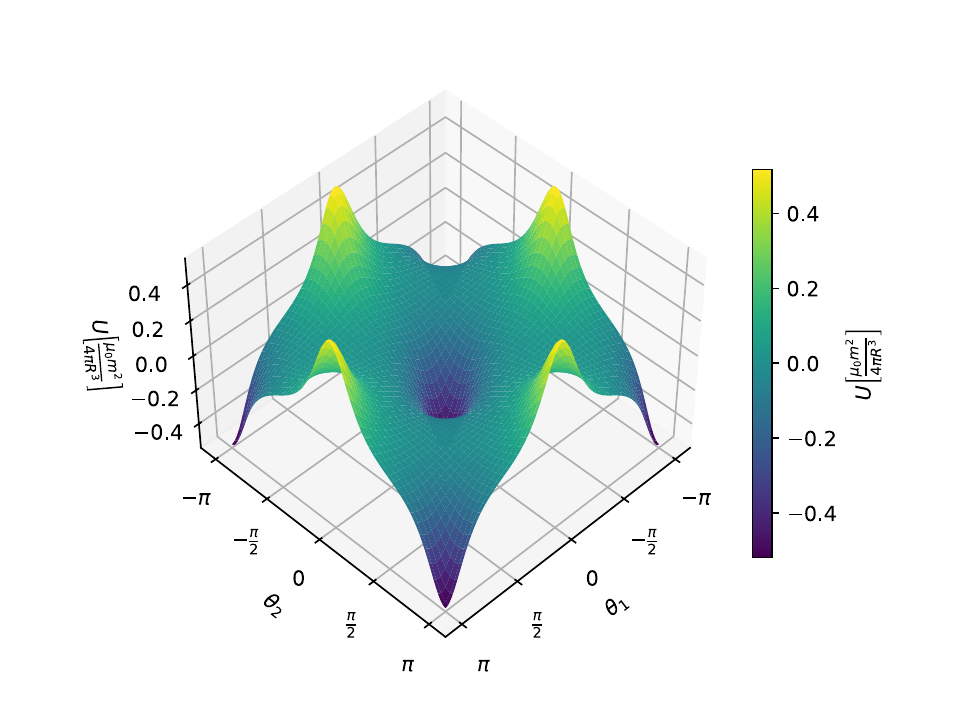}
			\caption{$N = 2$}
			\label{fig:potential_plot_3D_N_2}
		\end{subfigure}
		
		\begin{subfigure}[b]{0.45\textwidth}
			\centering
			\includegraphics[width=\textwidth]{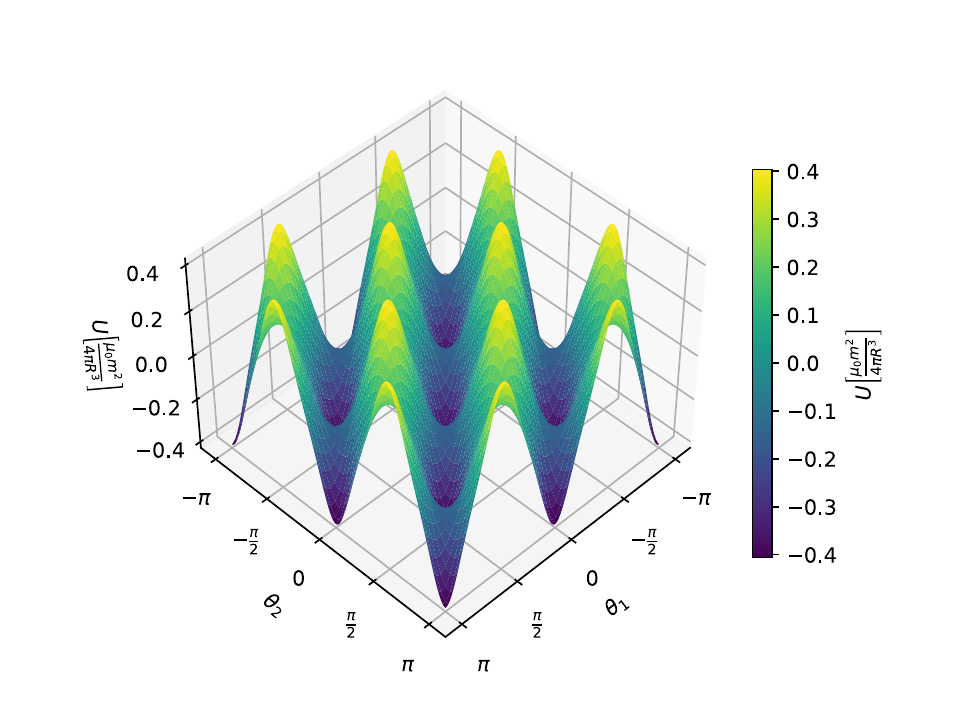}
			\caption{$N = 4$}
			\label{fig:potential_plot_3D_N_4}
		\end{subfigure}
		\hfill
		\begin{subfigure}[b]{0.45\textwidth}
			\centering
			\includegraphics[width=\textwidth]{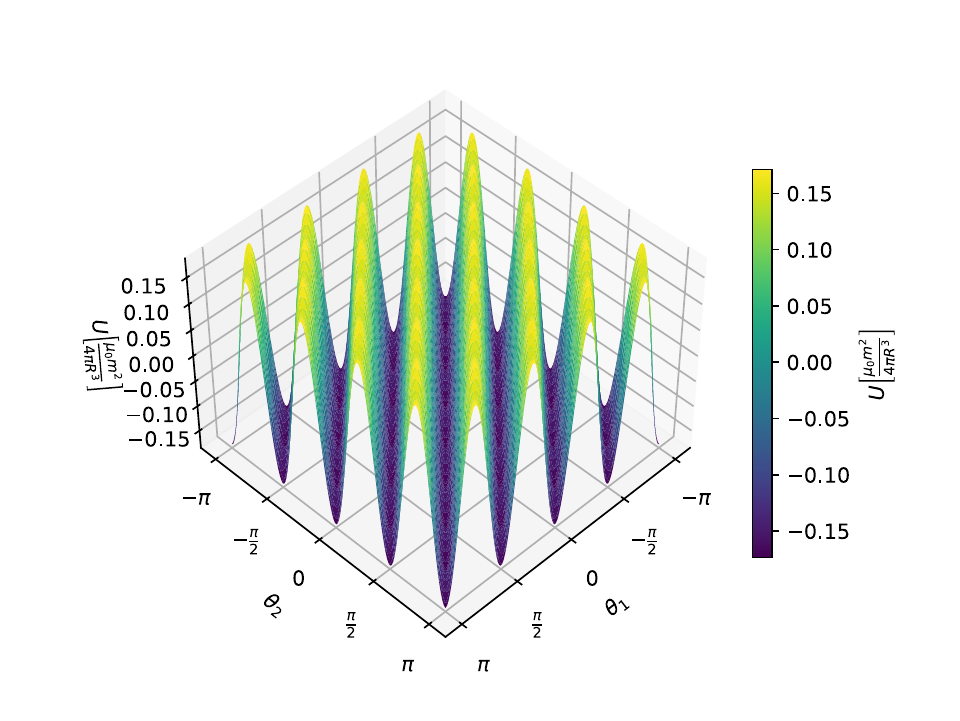}
			\caption{$N = 8$}
			\label{fig:potential_plot_3D_N_8}
		\end{subfigure}
		\caption{With the increase in the number of dipoles distributed on the gear, the higher saddle points become relatively closer to the potential barrier, making the particle less likely to "slip away."}
	\end{figure}
	
	\begin{figure}[h!]
		\centering
		\begin{subfigure}[b]{0.45\textwidth}
			\centering
			\includegraphics[width=\textwidth]{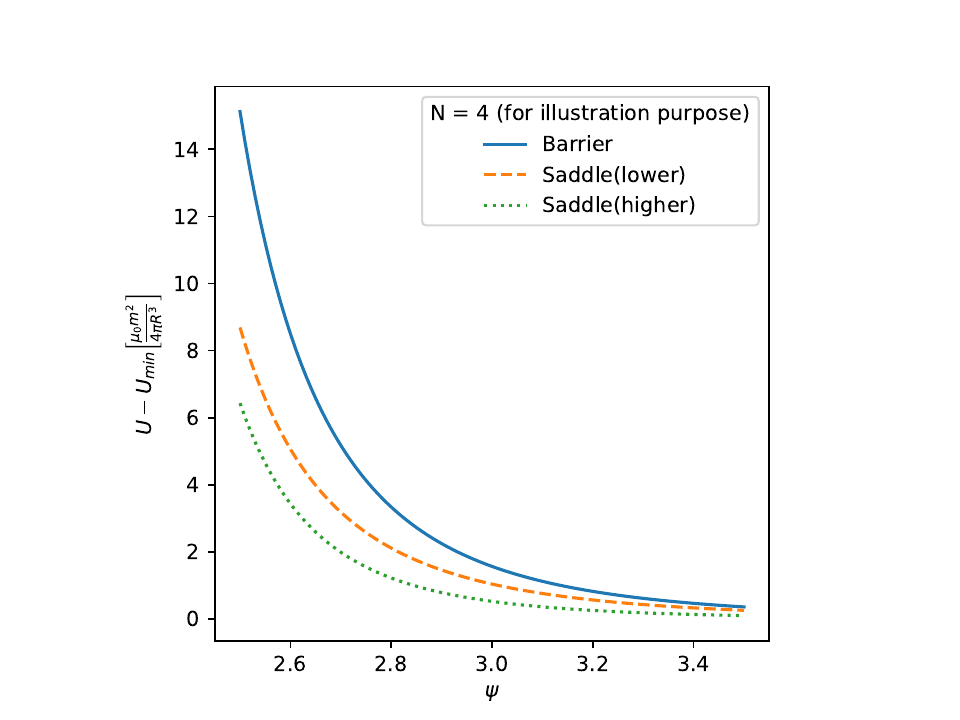}
			\caption{Characteristic potential values decrease as the two gears move away from each other.}
			\label{fig:diff-xi}
		\end{subfigure}
		\hfill
		\begin{subfigure}[b]{0.45\textwidth}
			\centering
			\includegraphics[width=\textwidth]{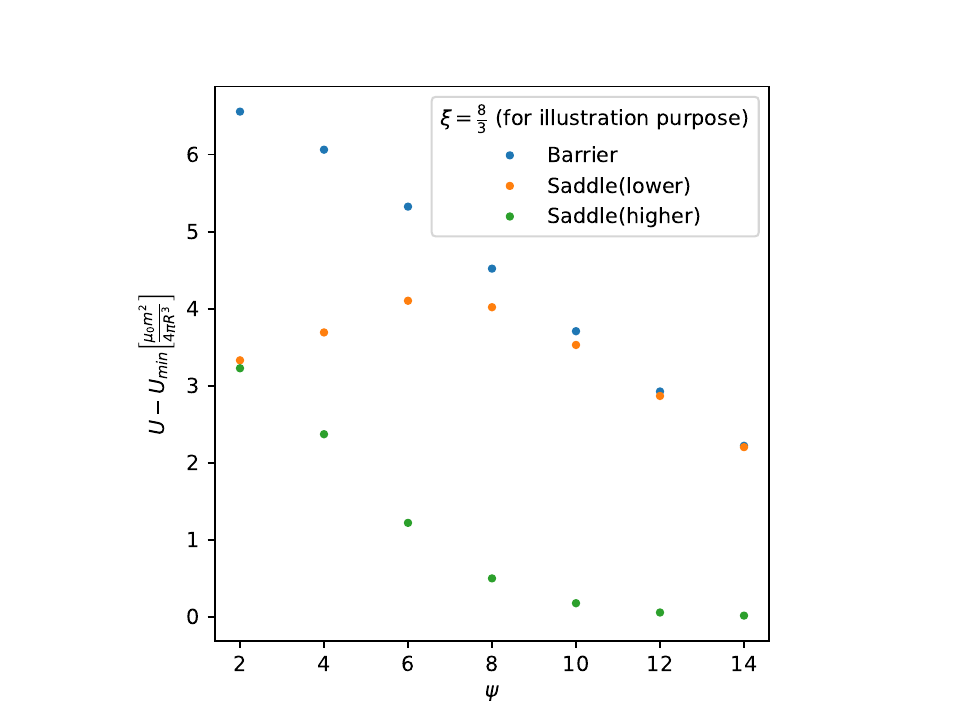}
			\caption{The more dipoles there are, the smaller the difference between the higher saddle point and the potential barrier.}
			\label{fig:diff-n}
		\end{subfigure}
		\caption{The relationship between characteristic potential values and the parameters $\xi$ and $N$.}
	\end{figure}

	\section{Stable Transmission and Forced Vibration}
	\subsection{Changing Stable Point}
\begin{figure}[h!]
		\centering
		\includegraphics[width=\textwidth]{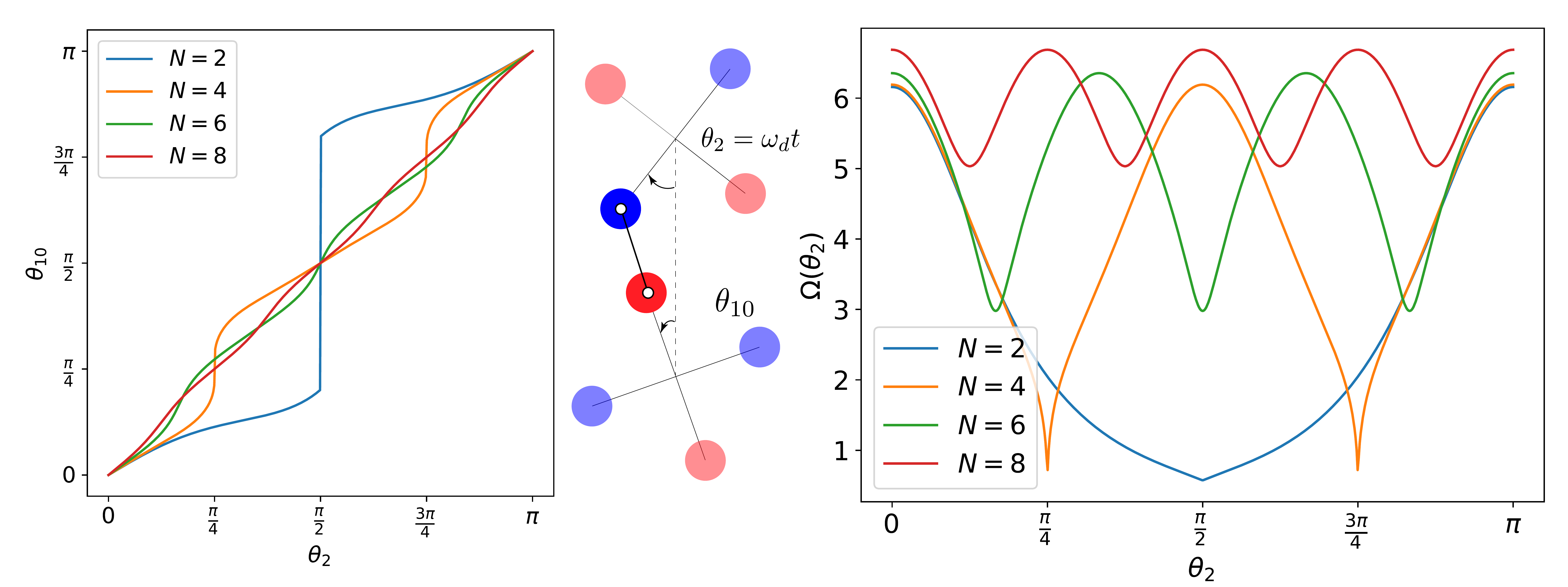}
		\caption{The left diagram indicates that the driven wheel does not rotate synchronously with the driving wheel. The right diagram indicates a sophiscated forced vibration system, as the resonance oscillation($\Omega$) keeps changing with $\theta_2$(or time) }
		\label{fig:balance}
	\end{figure}

	A more interesting phenomenon occurs when the second gear is rotated by a step motor at a constant speed. We are tend to imagine that as long as the second gear rotates slow enough, the first gear will always be able to rotate synchronously. This is based on a assumption that the first gear's energy is dissipative(i.e. energy loss due to friction etc.) . But it's quite intriguing that the two gear are moving not exactly synchronously. It has a alterating phase difference. We managed to calculate the equilibrium point for angle $\theta_1$ changes with second angle $\theta_2$.\ref{fig:balance}
	
	Because the potential energy term is inversely proportional to the cube of the distance between dipoles, the nearest pair dominates. This causes the relative positions of the master and slave wheels to deviate from the symmetric state most of the time. Whenever both rotate to an angle of $\frac{\pi}{2N}$, two pairs of magnetic dipoles simultaneously dominate, and near this point, the equilibrium position of the slave wheel changes rapidly—equivalent to a strong perturbation. Moreover, when the master wheel rotates at an angular frequency $f$, the slave wheel effectively experiences a periodic forced oscillation at N times the angular frequency f. Although the oscillation frequency increases by a factor of N, the shorter intervals between oscillations mean the slave wheel's rotation becomes more synchronized, and the amplitude of each disturbance decreases.
	
	\subsection{Experimental results for stable transmission}
	We built an experimental setup to verify our analytical results. We ordered several N42 sphere magnets, which exhibit stray fields that behave exactly like dipoles. To maintain the magnets' orientation along the axis of rotation of the spinners, we 3D printed different fidget spinner designs, each capable of holding 4 or 6 spheres. The center of the driven wheel contains a bearing, while the driving wheel features a hole for the stepper motor to fit into.
	\begin{table}[h]
		\centering
		\caption{parameters}
		\begin{tabular}{|c|c|c|c|c|c|}
			\hline
			$I$ [kg/m\(^2\)] & $m$ [A\(\cdot\)m\(^2\)] & $R$ [m] & $d$ [m] & N &$\omega_0 = \sqrt{\frac{\mu_0 m^2}{4 \pi R^{3} I}}$\\ \hline
			$3.45 \times 10^{-5}$ & 0.76 & 0.025 & 0.080 & 4 &10.4\\ \hline
		\end{tabular}
	\end{table}
	
	We set the step motor to various angular frequencies  $f$ and used a high-speed camera recording at up to 1000 fps to track the motion of the driven gear. We then plotted the relationship between the displacement angle $\theta_1-2\pi f t$  and time $t$. We notice the driven gear is oscillating while rotating along the driving gear. With the increase of the driving speed, the maximum oscillation angle increases until it fails to catch up with the driving gear.
	\begin{figure}[h!]
		\centering
		\includegraphics[width=\textwidth]{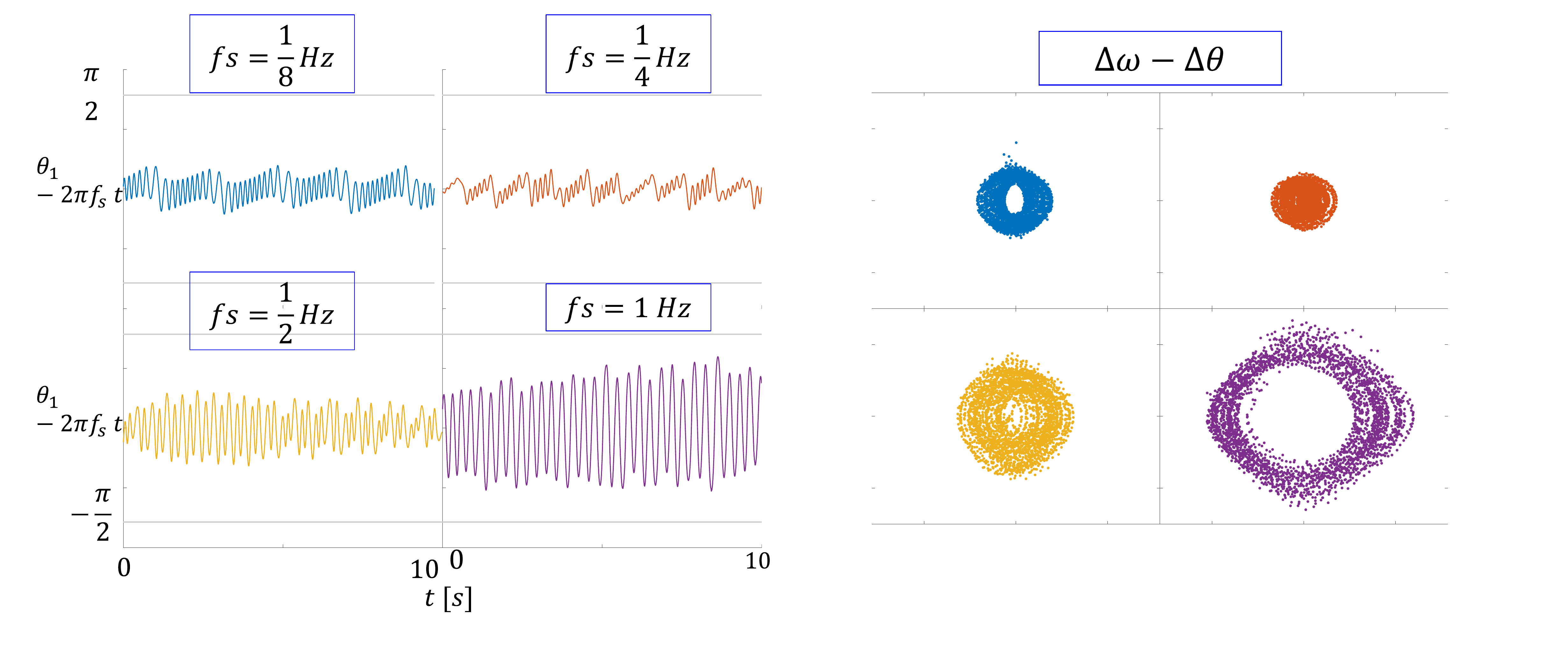}
		\caption{}
		\label{fig:exp}
	\end{figure}
	\section{Conclusion}
	This study illuminates the challenges faced by spur gears in achieving synchronous rotation. Unlike coaxial magnetic gears, which maintain stability at a fixed speed ratio, spur gear systems are inherently unstable under similar conditions.
	
	Firstly, the potential energy is inversely proportional to the cube of the distance between dipoles, causing only the nearest pairs to exert significant influence while those further away contribute minimally to the system's dynamics.
	
	Secondly, the dominant magnets on the driven gear tend to remain closer to the driving gear. This proximity disrupts the ideal condition where a fixed speed ratio is maintained, resulting in continuous oscillations of the angle difference. This instability is a critical factor that limits the effectiveness of spur gear systems in practical applications.

\end{document}